\documentclass[10pt,letterpaper]{article}
\usepackage{ol2}
\usepackage[draft]{hyperref}
\usepackage{amsmath}
\usepackage{graphicx}
\usepackage{floatrow}
\graphicspath{{pict/}{./}}
\usepackage[font={small}]{caption}

\usepackage{bm}%

\newcounter{Fig}

\newcommand{\be}{\begin{equation}}
\newcommand{\ee}{\end{equation}}

\begin{document}
\title{Elusive pure anapole excitation in homogenous spherical nanoparticles with radial anisotropy}
\author{Wei Liu,$^{1,*}$  Bing Lei,$^{1}$ Jianhua Shi,$^{1}$ Haojun Hu,$^{1}$ and Andrey E. Miroshnichenko$^2$}
\address{
$^1$College of Optoelectronic Science and Engineering, National University of Defense
Technology, Changsha, Hunan 410073, P. R. China.\\
$^2$Nonlinear Physics Centre, Research School of Physics and Engineering,  Australian National
University, Acton, ACT 0200, Australia.\\
$^*$Corresponding author: wei.liu.pku@gmail.com
}
\begin{abstract}
Copyright \copyright~~2015 Wei Liu \textit{et al.} This is an open access article distributed under the Creative Commons
Attribution License, which permits unrestricted use, distribution, and reproduction in any medium, provided the original work is
properly cited.
\end{abstract}
\begin{abstract}
For  homogenous isotropic dielectric nanospheres with incident plane waves, Cartesian electric and toroidal dipoles can be tunned to cancel each other in terms of far-field scattering, leading to the effective anopole excitation. At the same time however, other multipoles such as magnetic dipoles with comparable scattered power are simultanesouly excited, mixing with the anopole and leading to a non-negligible total scattering cross section. Here we show that for homogenous dielectric nanospheres, radial anisotropy can be employed to significantly suppress the other multipole excitation, which at the same time does not compromise the property of complete scattering cancallation between Cartesian electric and toroidal dipoles. This enables an elusive \textit{pure anopole} excitation within radially anisotropic dielectric nanospheres, which may shed new light to many scattering related fundamental researches and applications.
\end{abstract}
\section{Introduction}

Since the discovery by J. J. Thomson that electrons are constituents of atoms~\cite{Buchwald2004_Histories},  there had been a lot of searches for charge-current distributions in which no electromagnetic energy are radiated, which can be applied to explain the stability of atoms~\cite{Gbur2003_PO_Nonradiating}. This is due to the existing dogma since the establishment of Maxwell equations that accelerating charges would lose energy continuously through electromagnetic radiation, and consequently to prevent the collapse of atoms, the charges within them must be arranged into specific configurations. Such special charge-current configurations are also termed as non-radiating sources~\cite{Gbur2003_PO_Nonradiating,jackson1962classical}. Those non-radiating sources are deeply related to the concepts of anapoles in quantum field theory, which have their roots in the law of parity unconservation~\cite{Zel'Dovich1958_JETP}, and can be used as possible explanations for the existence and features of neutrinos and cold dark matter~\cite {Dubovik1998_IJMPA_toroid,Ho2013_PLB_anapole}.

For time-varying oscillating charge-current distributions, the dynamic version of anapoles is highly related to the so called toroidal multipoles, which recently have attracted enormous attention~\cite{Kaelberer2010_Science,dong2012toroidal,Huang2012_OE,Ogut2012_NL,Fedotov2013_SR,Fan2013_lowloss,savinov2014toroidal,Xiao2014_AP_core,Basharin2014_arXiv,miroshnichenko2014seeing,Liu2014_arxiv,Bakker2015_NL_Magnetic,
Liu2015_OL_invisible,Liu2015_arXiv_Efficient}. As independent terms in the multipole expansion of electromagnetic fields besides conventional cartesian magnetic and electric multipoles~\cite{jackson1962classical}, toroidal multipoles correspond to significantly different near-field charge-current distributions and will inevitably become significant when the operating spectral regime goes beyond the electrostatic region~\cite{Radescu2002_PRE,Liu2015_arXiv_Efficient,Corbaton2015_arXiv_Exact}.  In the far field however, the toroidal multipoles have identical scattering patterns compared with their conventional Cartesian electric and magnetic counterparts, and that is to say,  they are indistinguishable in terms of far-field scattering~\cite{Radescu2002_PRE,Kaelberer2010_Science,Chen2011_NP531,savinov2014toroidal}. This at the same time also means that when properly engineered, the contributions of conventional and toroidal multipoles to the scattered field can be tuned to the same amplitude and out of phase, which leads to the far-field scattering cancellation and thus generate the so called electrodynamic anapoles~\cite{Dubovik1990_PR,Radescu2002_PRE,Fedotov2013_SR,Basharin2014_arXiv,miroshnichenko2014seeing,Liu2014_arxiv,Liu2015_OL_invisible}. The simplest case of such anapoles is induced by the destructive cancellation of Cartesian electric dipole and toroidal dipole, which has been demonstrated firstly in composite carefully engineered structures consisting of particle clusters~\cite{Fedotov2013_SR,Basharin2014_arXiv}.  Recently it has  also  been theoretically proposed and  experimentally demonstrated in surprise that the anapole can be effectively excited  even in individual homogenous dielectric nanoparticles~\cite{miroshnichenko2014seeing}. However, the problem of the above demonstration in simple structures is that besides the anapole excitation, other multipoles (such as magnetic dipoles) of similar magnitudes in terms of scattered power are also excited and mixed with the excited electric and toroidal dipoles~\cite{miroshnichenko2014seeing,Liu2014_arxiv}. Such problems can be conquered within single core-shell nanoparticles of fundamental shapes (spherical or cylindrical) where the extra multipole excitations can be significantly suppressed, enabling elusive \textit{pure anapole} excitations~\cite{Liu2014_arxiv,Liu2015_OL_invisible}.

Here we propose an alternative approach and demonstrate that elusive \textit{pure anapole} can be observed in homogenous radially anisotropic dielectric spherical nanoparticles under incident plane waves. It is shown that when the radial anisotropy of material permittivity is employed, the Cartesian electric and toroidal dipoles can be tuned to the spectral positions where other multipoles are significantly suppressed, without compromising the feature of complete destructive interference between them. Since Cartesian electric and toroidal dipoles can totally cancel each other in terms of far-field scattering and at the same time the excitation of other multipoles is negligible, it is proved that effectively \textit{pure anapole} has been obtained. The anapole excitation is highly related to the exotic scattering phenomena of invisibility and possible applications of non-invasive sensing and detections,  which may pave the way for many further scattering related fundamental researches and applications.

\section{Methods and Expressions}

\subsection{Multipole Expansions with toroidal components}

For any charge-current distribution, when the geometric size of the source is far smaller than the effective wavelength (the retardation effect within the source can be neglected), the quasi-static approximation can be applied and then the radiated fields of the source can be expanded into two sets of conventional Cartesian electric and magnetic multipoles~\cite{jackson1962classical,Radescu2002_PRE}. The lowest order terms of the two sets of multipoles are electric and magnetic dipoles, which can be expressed respectively as [throughout the paper we adopt the $\rm exp(-i\omega t+i\textbf{k}\cdot\textbf{r})$ notation for electromagnetic waves]:
\begin{equation}
\label{ED_TD}
\textbf{P}= {1 \over { - i\omega }}\int d^3r\textbf{J}(\textbf{r}), ~~\textbf{M} = {1 \over {2c}}\int {d^3 r\left[ {\textbf{r }\times \textbf{J}(\textbf{r})} \right]^{}},
\end{equation}
where $c$ is the speed of light, $\omega$ is the angular frequency and $\textbf{J}$ is the current. When the retardation effect within the source cannot be neglected (the size of the source gets larger and/or the effective wavelength gets smaller), the two sets of multipoles are not complete and thus other multipoles have to be included as correcting terms~\cite{Radescu2002_PRE,Liu2015_arXiv_Efficient,Corbaton2015_arXiv_Exact}. The lowest order correcting multipoles are toroidal multipoles, with the toroidal dipole expressed as as~\cite{Dubovik1990_PR,Radescu2002_PRE,Kaelberer2010_Science,Chen2011_NP531}:
\begin{equation}
\label{TD}
\textbf{T}= {1 \over {10c}}\int d^3r[(\textbf{r} \cdot \textbf{J}(\textbf{r}))\textbf{r} - 2r^2 \textbf{J}].
\end{equation}

The radiated power of  $\textbf{P}$,  $\textbf{T}$  and  $\textbf{M}$ are respectively:
\begin{eqnarray}
\rm W_{\rm \textbf{P}}  = {{\mu _0 \omega ^4 } \over {12\pi c}}\left| \textbf{P} \right|^2,\label{scattered_p}\\
\rm W_{\rm \textbf{T}}  = {{\mu _0 \omega ^4 k^2 } \over {12\pi c}}\left| \textbf{T}\right|^2,\label{scattered_t}\\
\rm W_{\rm \textbf{M}}  = {{\mu _0 \omega ^4 } \over {12\pi c}}\left| \textbf{M} \right|^2,\label{scattered_m}
\end{eqnarray}
where $\mu _0$ is the vacuum permeability and $k$ is the angular wave number.  We emphasize that the expressions above are for individual multipoles only and a simple sum of them does not lead to the total scattered power as there is interference between them~\cite{Radescu2002_PRE,Kaelberer2010_Science,Chen2011_NP531} [see also Eq.~(\ref{Scattering_dipolar}) below]. It is also worth mentioning that the incorporation of toroidal multipoles does not make the expansion complete~\cite{Radescu2002_PRE,Liu2015_arXiv_Efficient,Corbaton2015_arXiv_Exact}. When the operating spectral region is far beyond the quasi-static regime, besides toroidal multipoles, other higher order correcting multipoles have to be also included~\cite{Radescu2002_PRE,Liu2015_arXiv_Efficient,Corbaton2015_arXiv_Exact}.

For scattering particles, based on the charge conservation relation, the current (displacement current) distribution is
\begin{equation}
\label{current}
\textbf{J}(\textbf{r}) = -i\omega \varepsilon _0 [\varepsilon(\textbf{r})-1]\textbf{E}(\textbf{r}),
\end{equation}
where $\varepsilon _0$ is the vacuum permittivity; $\varepsilon(\textbf{r})$ is the relative permittivity, which is a scalar for isotropic particles or tensor for anisotropic particles. With the current distribution shown in Eq.(\ref{current}),  all the Cartesian multipoles including $\textbf{P}$,  $\textbf{T}$  and  $\textbf{M}$ and their scattered power can be directly calculated [see Eqs.(\ref{ED_TD})-(\ref{scattered_m})].

\subsection{Electric and magnetic multipoles deduced from vector spherical harmonics of Mie scattering particles}

For spherical (both isotropic and radially anisotropic) particles with incident plane waves of electric field $\textbf{E}_0$ and magnetic field $\textbf{H}_0$, both the fields inside the particle and the scattered fields can be expressed in transverse electric and transverse magnetic (in terms of electric field distribution)vector spherical harmonics~\cite{Bohren1983_book,Bohren1983_book,Qiu2008_JOSAA,Qiu2010_IPOR_Light,Liu2015_OE_Ultra,Liu2015_arXiv_Efficient}. On one hand, according to Eqs.(\ref{ED_TD})-(\ref{current}),  all the Cartesian multipoles including $\textbf{P}$,  $\textbf{T}$  and  $\textbf{M}$ and their scattered power can be analytically expressed and numerically calculated. On the other hand, based on the far-field equivalence, the scattered fields can be viewed as the radiated fields of a combination of conventional Cartesian electric and magnetic multipoles (which correspond to transverse magnetic and transverse electric vector spherical harmonics respectively). As a result the far-field deduced electric dipole $\textbf{P}(a_1)$ and magnetic dipole $\textbf{M}(b_1)$  can be expressed respectively as~\cite{Bohren1983_book,Doyle1989_optical,miroshnichenko2014seeing,Liu2012_ACSNANO,Liu2012_PRB081407,Liu2014_arxiv,Wheeler2006_PRB,Liu2015_OE_Ultra,Liu2015_arXiv_Efficient}:
\begin{equation}
\label{conventioanl_dipole}
\textbf{P}(a_1)= \varepsilon _0 {{6\pi ia_1 } \over {k^3 }}\textbf{E}_0,~~\textbf{M}(b_1)= \mu _0 {{6\pi ib_1 } \over {k^3 }}\textbf{H}_0.
\end{equation}
where $a_1$ and $b_1$ are the Mie scattering coefficients of the lowest order.  The total scattered power is ~\cite{Bohren1983_book,Qiu2008_JOSAA,Qiu2010_IPOR_Light,Liu2015_OE_Ultra}:

\begin{equation}
\label{all}
{\rm W_{\rm All}} ={{\pi \left| {E_0 } \right|^2 } \over {k\omega \mu _0 }}\sum\limits_{m = 1}^\infty  {(2m + 1)({{\left| {{a_m}} \right|}^2}}  + {\left| {{b_m}} \right|^2}).
\end{equation}
where $a_m$ and $b_m$ are the Mie scattering coefficients of order $m$ and $E_m  = i^m E_0 {{2m + 1} \over {m(m + 1)}}$.  It is clear that the scattered power of $\textbf{P}(a_1)$ and $\textbf{M}(b_1)$ is respectively:
\begin{equation}
\label{w_ab}
{\rm W_{\rm \textbf{P}(a_1)}} ={{3\pi \left| {E_0 } \right|^2 } \over {k\omega \mu _0 }}\left|a_1\right|^2,~{\rm W_{\rm \textbf{M}(b_1)}} ={{3\pi \left| {E_0 } \right|^2 } \over {k\omega \mu _0 }}\left|b_1\right|^2.
\end{equation}

\subsection{Anapole excitation induced by destructive interference between $\textbf{P}$ and $\textbf{T}$}
As is discussed above, $\textbf{P}(a_1)$  shown in Eq.(~\ref{conventioanl_dipole})  has been obtained based on the complete vector spherical harmonic expansion and
thus has included all the transverse magnetic dipolar scattering components~\cite{Doyle1989_optical,miroshnichenko2014seeing,Liu2012_ACSNANO,Liu2012_PRB081407,Liu2014_arxiv,Wheeler2006_PRB,Liu2015_OE_Ultra,Liu2015_arXiv_Efficient}. While the conventional Cartesian electric dipole P shown in Eq.~(\ref{ED_TD}) is calculated through current integration and is only the lowest order expansion of $\textbf{P}(a_1)$~\cite{Radescu2002_PRE,Liu2015_arXiv_Efficient,Corbaton2015_arXiv_Exact}. For operating spectral region beyond the quasi-static regime, the next order correcting term, which is the toroidal dipole $\textbf{T}$  shown in Eq.(~\ref{conventioanl_dipole}), should also be included. When higher order correcting terms beyond toroidal terms  can be neglected,
the scattered dipolar transverse magnetic field can be expressed as as~\cite{Radescu2002_PRE,Chen2011_NP531,miroshnichenko2014seeing,Liu2014_arxiv}:
\begin{equation}
\label{Scattering_dipolar}
\textbf{E}_{\rm rad}  ={{k^2 \exp (ikr)} \over {4\pi r\varepsilon _0 }}\textbf{n} \times \textbf{P}(a_1) \times \textbf{n}\approx {{k^2 \exp (ikr)} \over {4\pi r\varepsilon _0 }}\textbf{n} \times (\textbf{P} + ik\textbf{T}) \times \textbf{n},
\end{equation}
where $\textbf{n}$ is the unit vector along $\textbf{r}$. The existence of the anapole lies in $\textbf{P}(a_1)\approx\textbf{P} + ik\textbf{T}=0$ which can be satisfied when:
\begin{equation}
\label{anapole}
 \textbf{P}=-ik\textbf{T}
\end{equation}

It is worth mentioning that when Eq.(~\ref{anapole}) is satisfied, only the transverse magnetic dipolar scattering will be vanishing and consequently the anapole excited (consisting of destructively interfering  $\textbf{P}$ and $\textbf{T}$) can be mixed with other excited multipoles, such as magnetic dipoles, electric quadrupoles and so on~\cite{miroshnichenko2014seeing,Liu2014_arxiv,Liu2015_OL_invisible}.

\begin{figure}
\centerline{\includegraphics[width=8.5cm]{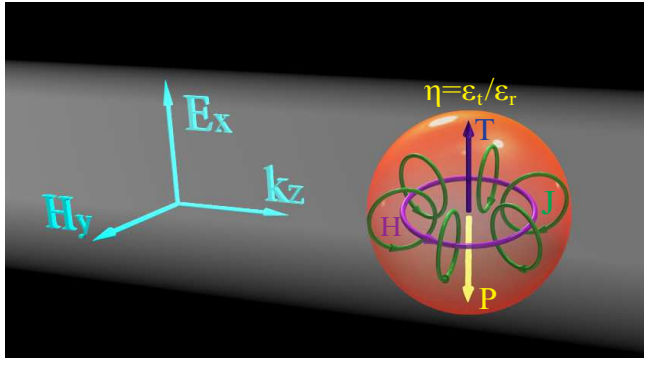}}\caption{\small  Schematic illustration of anopole excitation in a homogenous particle with plane wave incidence. The radius of the particle is R,  and the particle is isotropic with refractive index $n$, or is radially anisotropic with transverse permittivity $\varepsilon_t=n^2$ and radial permittivity $\varepsilon_r$ (the anisotropy parameter is expressed as: $\eta=\varepsilon_t/\varepsilon_r$). Within the particle, both $\textbf{T}$ and $\textbf{P}$ are excited, out of phase and of the same magnitude with regard to the scattered power. At the same time, the typical current $\textbf{J}$ and magnetic field $\textbf{H}$ distributions for $\textbf{T}$ only have been shown.}
\label{fig1}
\end{figure}

\begin{figure*}
\centerline{\includegraphics[width=15cm]{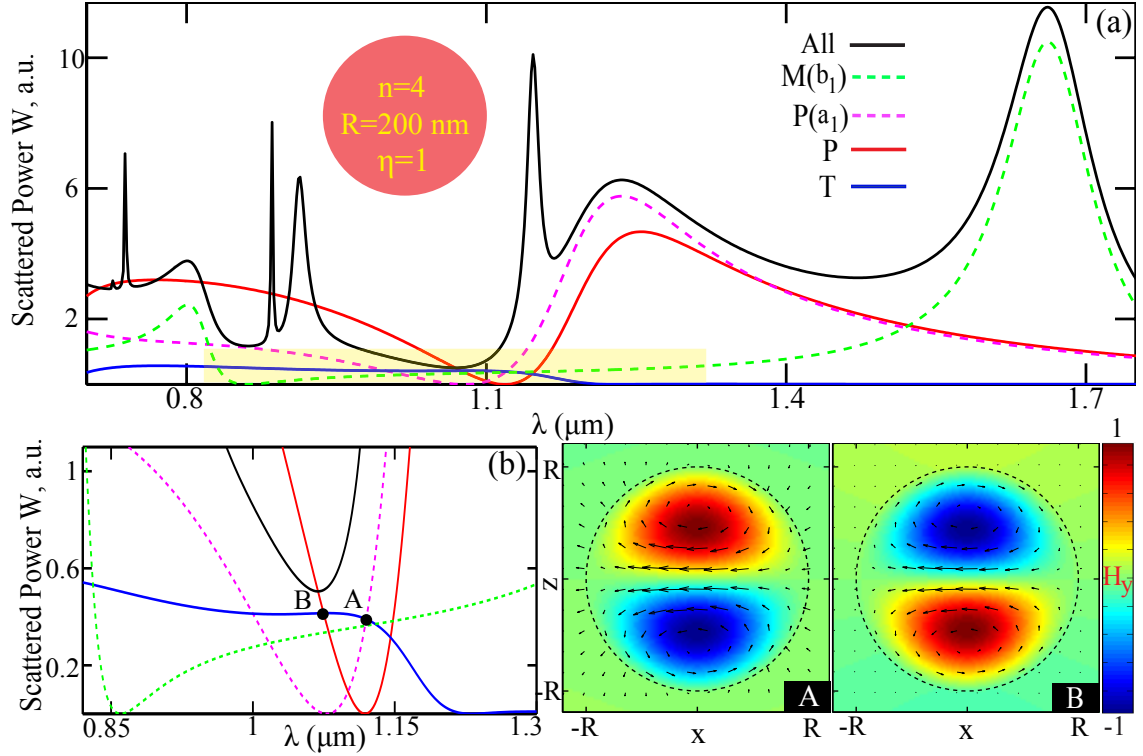}}\caption{\small (a) Scattered power spectra for an isotropic ($\eta=1$) dielectric ($n=4$) sphere of $R=200$~nm  (inset). Besides the total scattered power (All, black curve), the contributions from current-integrated electric dipole ($\textbf{P}$, red curve), toroidal dipole ($\textbf{T}$, blue curve), and those from the far-field deduced electric dipole [$\rm{\textbf{P}(a_1)}$, dashed magenta curve] and magnetic dipole [$\rm{\textbf{M}(b_1)}$, dashed green curve] are shown. The shaded area is of interest and shown in detail in (b). Two points of different $\lambda$ are indicated by black dots (A: $\lambda=1118$~nm, B: $\lambda=1075$~nm). The corresponding near fields on the $x-z$ plane of $y=0$ are shown respectively in (A)-(B).  The distributions for both the normalized $\textbf{H}_y$ (color-plots) and displacement field $\textbf{D}=\varepsilon\textbf{E}$ on the $x-z$ plane (vector-plots) are shown (the dashed black lines denote the boundaries of the particles), as is also the case for Fig.~\ref{fig4}.}
\label{fig2}
\end{figure*}

\begin{figure}
\centerline{\includegraphics[width=7.5cm]{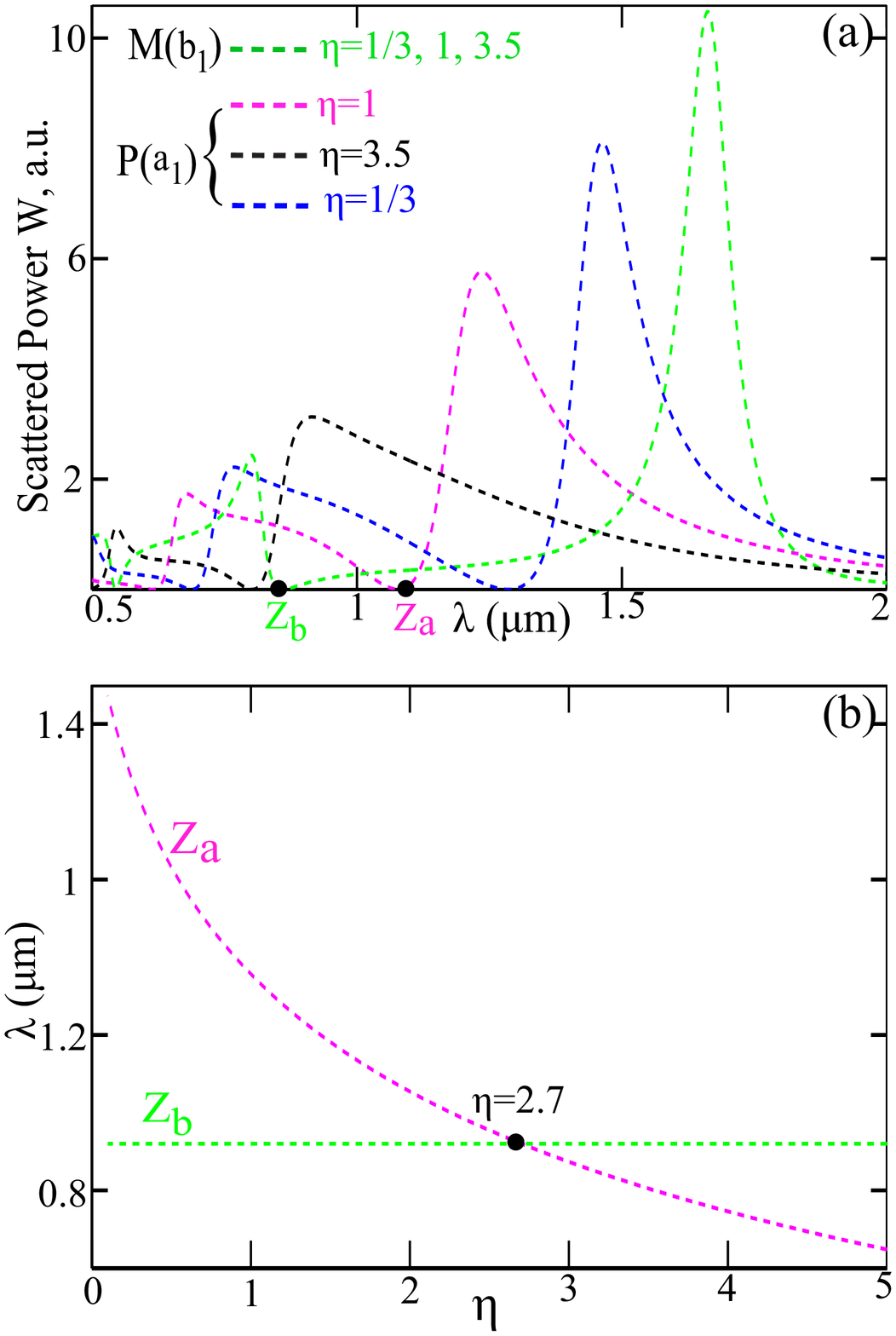}}\caption{\small (a) Scattering spectra of radially anisotropic dielectric nanosphere (transverse permittivity is $\varepsilon_t=4^2$ and $R=200$~nm) showing only contributions from $\rm{\textbf{P}(a_1)}$ and $\rm{\textbf{M}(b_1)}$.  The lowest order minimum scattering points of $\rm{\textbf{P}(a_1)}$ and $\rm{\textbf{M}(b_1)}$  are pinpointed by  circular dots  $\rm{\textbf{Z}_a}$  and $\rm{\textbf{Z}_b}$ respectively. (b) Spectral dependence of the minimum scattering points on $\eta$ is shown in (b) and  $\eta=2.7$ is the overlapping point of $\rm{\textbf{Z}_a}$  and $\rm{\textbf{Z}_b}$.}
\label{fig3}
\end{figure}

\section{Results and Discussions}

In Fig.~\ref{fig1} we provide a schematic illustration of anopole excitation in a homogenous particle with plane wave incidence. The plane wave is polarized along $x$ direction (in terms of electric field) and propagating along $z$ direction. The radius of the particle is R, and the particle is isotropic with refractive index $n$, or is radially anisotropic with transverse (perpendicular to the radial direction) permittivity $\varepsilon_t=n^2$ and radial permittivity $\varepsilon_r$, with the anisotropy parameter defined as: $\eta=\varepsilon_t/\varepsilon_r$. Both $\textbf{T}$ and $\textbf{P}$ are excited, out of phase and of the same magnitude with regard to the scattered power, indicating excitation of the anapole [Eqs.~(\ref{Scattering_dipolar})-(\ref{anapole})]. Also the typical current $\textbf{J}$ and magnetic field $\textbf{H}$ distributions for $\textbf{T}$ only have been shown.

\subsection{Mixed excitation of anapole and other multipoles within homogeneous isotropic dielectric nanospheres}

At the beginning, we investigate the scattering of homogeneous isotropic ($\eta=1$) dielectric nanospheres with plane wave incidence with $R=200$~nm and $n=4$ [see the inset of Fig.~\ref{fig2}(a)]. The scattered power spectra are in Fig.~\ref{fig2}(a):  the contributions from far field deduced electric and magnetic dipoles $\textbf{P}(a_1)$ and $\textbf{M}(b_1)$, conventional Cartesian electric and toroidal dipoles $\textbf{P}$ and $\textbf{T}$, and the total scattered power are shown. The shaded region in Fig.~\ref{fig2}(a) is the area of interest, and thus is zoomed in and shown in Fig.~\ref{fig2}(b), where two points A and B are pinpointed and indicated by black dots. At point A of $\lambda=1118$~nm, for the  transverse magnetic dipolar components, there is only toroidal dipole $\textbf{T}$ excitation [$\textbf{P}(a_1) \approx ik\textbf{T}$], and the conventional Cartesian electric dipole $\textbf{P}$ is completely suppressed; At point B of $\lambda=1075$~nm, there is equal out of phase $\textbf{T}$ and $\textbf{P}$  excitation in terms of scattering power, leading to an effective anapole excitation defined in Eqs.~(\ref{Scattering_dipolar})-(\ref{anapole})~\cite{miroshnichenko2014seeing,Liu2014_arxiv,Liu2015_OL_invisible,Liu2015_arXiv_Efficient}. The corresponding near-field distributions are shown in Fig.~\ref{fig2}(A-B): the vector-plots correspond to the displacement field $\textbf{D}=\varepsilon\textbf{E}$ on the $x-z$ plane and the color-plots correspond to the normalized magnetic fields along $y$ direction ($\textbf{H}_y$). We note here that for better visibility we plot only transverse magnetic dipolar fields and as is shown: (1) At point A shown in Fig.~\ref{fig2}(A), the near field  exhibits a typical toroidal dipolar distribution, as is schematically shown in  Fig.~\ref{fig1} (the magnetic field is confined within the particle and there are accompanying oppositely circulating current loops); (2) Point B shown Fig.~\ref{fig2}(B) corresponds to an anapole excitation, and at this point the fields outside the particle are vanishing, confirming the negligible overall dipolar scattering and thus effective anapole excitation.

According to Fig.~\ref{fig2}(b), at Point B besides the anapole excitation, there is the coexistence of other multipoles, of which the far field deduced magnetic dipole $\textbf{M}(b_1)$ is the most significant one. At the same time, the scattered power of $\textbf{M}(b_1)$ is close to that of both $\textbf{P}$ and $\textbf{T}$, rendering the anapole excitation of a hybrid and mixed nature. Basically to obtain the sole excitation of a pure anapole, the suppression of $\textbf{M}(b_1)$ is required.
\begin{figure}
\centerline{\includegraphics[width=9cm]{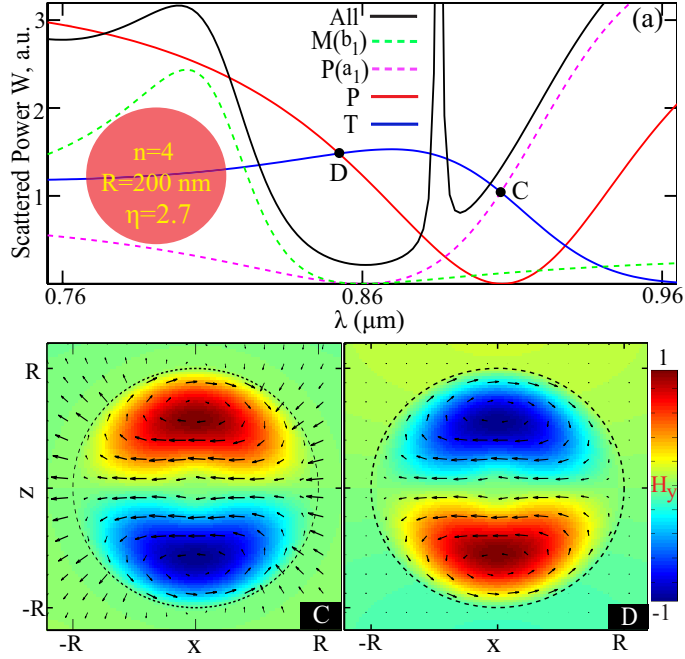}}\caption{\small (a) Scattered power spectra for an anisotropic ($\eta=2.7$) sphere  of $R=200$~nm and $\varepsilon_t=4^2$ (inset). The total scattered power together with the contributions from $\textbf{P}$, $\textbf{T}$, $\rm{\textbf{P}(a_1)}$, and $\rm{\textbf{M}(b_1)}$ are shown. Two points C and D of $\lambda=907$~nm and $852$~nm respectively are selected and the corresponding near fields are shown in (C)-(D).}
\label{fig4}
\end{figure}

\subsection{Overlapping the minimum scattering points of $\rm{\textbf{P}(a_1)}$ and $\rm{\textbf{M}(b_1)}$ within radially anisotropic nanospheres}

It is recently demonstrated that radial anisotropy can be employed to tune spectrally the far field deduced electric dipoles $\rm{\textbf{P}(a_1)}$ and thus the anapoles excited~\cite{Liu2015_OE_Ultra,Liu2015_arXiv_Efficient}. In contrast when the transverse permittivity $\varepsilon_t=n^2$ is fixed, the spectra of far field deduced magnetic dipoles $\rm{\textbf{M}(b_1)}$ are independent of the anisotropy parameter due to its transverse electric nature~\cite{Liu2015_OE_Ultra}.
In Fig.~\ref{fig3}(a) we show the scattering spectra of radially anisotropic dielectric nanosphere (transverse permittivity is fixed at $\varepsilon_t=4^2$ and
$R=200$~nm). Only the contributions from $\rm{\textbf{P}(a_1)}$ and $\rm{\textbf{M}(b_1)}$ are shown. It is clear that $\rm{\textbf{P}(a_1)}$ would blueshift/redshift with larger/smaller $\eta$, while $\rm{\textbf{M}(b_1)}$ is stable and independent of $\eta$.  We indicate the first (counted from large wavelength) minimum scattering points of $\rm{\textbf{P}(a_1)}$ and $\rm{\textbf{M}(b_1)}$  by  circular dots  $\rm{\textbf{Z}_a}$ (which corresponds to the anapole excitation~\cite{Liu2015_arXiv_Efficient}) and $\rm{\textbf{Z}_b}$ respectively in Fig.~\ref{fig3}(a), and their spectra dependence on $\eta$ is shown Fig.~\ref{fig3}(b).  According to Fig.~\ref{fig3}(b), besides the confirmation of the spectral tunability of $\rm{\textbf{P}(a_1)}$ and stability of $\rm{\textbf{M}(b_1)}$, it is further shown that at $\eta=2.7$ the minimum scattering points of both $\rm{\textbf{P}(a_1)}$ and $\rm{\textbf{M}(b_1)}$ overlap. This means the anapole excitation is achieved with simultaneous magnetic dipole suppression.

\subsection{Pure anapole excitation within radially anisotropic nanospheres}

To further show the significance of the overlapping of the minimum scattering points of $\rm{\textbf{P}(a_1)}$ and $\rm{\textbf{M}(b_1)}$ and the sole anapole excitation, we show in Fig.~\ref{fig4}(a) the spectra at $\eta=2.7$ for the radially anisotropic nanosphere with $\varepsilon_t=4^2$ and $R=200$~nm [see the inset of Fig.~\ref{fig4}(a)]. Similar to  Fig.~\ref{fig2}(a), contributions from $\textbf{P}(a_1)$ , $\textbf{M}(b_1)$, $\textbf{P}$, $\textbf{T}$ together with the total scattered power are shown.  Two points C ($\lambda=907$~nm) and D ($\lambda=852$~nm) are indicated by black dots [the corresponding near fields are shown in Fig.~\ref{fig4}(C)-Fig.~\ref{fig4}(D) respectively], whose counterparts (which are similar in terms of both scattering spectra and near-field distributions) are respectively points A and B in Fig.~\ref{fig2}(a): at point C there is dominant $\textbf{T}$ excitation [with magnetic field confined within the particle and oppositely circulating currents, see Fig.~\ref{fig4}(C)] and point D corresponds to the excited anapole [with almost zero fields everywhere outside the particle, see Fig.~\ref{fig4}(D)]. By comparing point B [see Fig.~\ref{fig2}] and D [see Fig.~\ref{fig4}], the significance of introducing radial anisotropy is clear: (1) The anapole is excited at the minimum scattering point of $\rm{\textbf{M}(b_1)}$ and at the same time besides the significant suppression of  $\rm{\textbf{M}(b_1)}$, other multipoles including  the electric and magnetic quadrupoles are also negligible (not shown here but is clear according to the total scattering), leading to vanishing total scattering; (2) Though completely cancel each other, the scattering of $\textbf{P}$ and $\textbf{T}$ is an order of magnitude larger than all the other multipoles, showing a very efficient anapole excitation. The two points above are sufficient to justify the claim of elusive pure anapole excitation within radially anisotropic nanospheres.

\section{Conclusions and Outlook}

To conclude, in this work we show the elusive excitation of \textit{pure anapoles} in homogenous nanoparticles, which has been realized through employing radial anisotropy of the material permittivity to tune the anapole excited to overlap spectrally with the minimum scattering position of magnetic dipoles. At the overlapping position, Cartesian electric and toroidal dipoles can totally cancel the scattering of each other in the far field and at the same time the excitation of other multipoles (including but not limited to magnetic dipoles) is negligible,  proving that the pure anapole has been achieved.

We emphasize that the anapole excited has been induced by the interferences between Cartesian electric and toroidal dipoles, which is featured by null scattering in terms of only transverse magnetic dipolar scattering. In a similar way, anapoles characterized by null transverse electric dipolar scattering, or both transverse electric and magnetic null scattering of higher orders can be achieved. The investigations into scattering nanoparticles from the new perspective of anapole and toroidal multipole excitations can also be conducted for other non-spherical structures where numerical methods can be applied to calculate the fields, currents, and thus multipoles excited.  Also in this work we have confined our study to correcting terms up to toroidal multipoles~\cite{Dubovik1990_PR,Radescu2002_PRE,Corbaton2015_arXiv_On,Liu2015_arXiv_Efficient}. When the operating spectral region is far beyond the quasi-static regime, higher order correcting terms than toroidal multipoles (in terms of $kR$) will arise inevitably~\cite{Dubovik1990_PR,Radescu2002_PRE,Corbaton2015_arXiv_On,Liu2015_arXiv_Efficient}, and as a result it is expected that anapoles induced by the inferencing between more than two multipoles can be obtained.

Here in this work we show only the demonstration and tuning of anapoles based on radial anisotropy, and other effects, such as nonlinear and gain effects, can also be applied for both near and far-field scattering manipulations based on anapole control. The investigations conducted into the fundamental Mie scattering problem from the new research angle of anapole excitation and tuning is of great significance for not only deepening our understanding of the basic scattering physics, but also for quite a few scattering related applications, such as far-field scattering pattern shaping~\cite{Liu2014_CPB,Liu2012_ACSNANO,Liu2014_ultradirectional,Liu2015_OE_Ultra}, non-invasive detection and sensing~\cite{Alu2009_PRL} and so on. The anapole mode exhibits a number of nontrivial topological properties, which may render new opportunities for manipulating electromagnetic waves in an unprecedented way when combined with the emerging fields of topological and low dimensional photonics. Moreover, the investigations into anapoles should not be confined to the field of electrodynamics, and similar studies can certainly be conducted in other fields such as acoustics, fluid dynamics, quantum field theory and so on.

\section*{Conflict of Interests}
The authors declare that there is no conflict of interests regarding the publication of this paper.

\section*{Acknowledgments}
We thank Jianfa Zhang  and Yuri S. Kivshar for valuable discussions, and acknowledge the financial support from the National Natural Science Foundation of China (Grant numbers: $11404403$ and $61205141$), the Australian Research Council (Grant number: FT110100037) and the Basic Research Scheme of College of Optoelectronic Science and Engineering, National University of Defense Technology.


\end{document}